%% file: paper.tex
\documentclass[fleqn,11pt]{wlscirep}

\usepackage[utf8]{inputenc}%
\usepackage[T1]{fontenc}%
\usepackage{graphicx}%
\usepackage{multirow}%
\usepackage{amsmath,amssymb,amsfonts}%
\usepackage{amsthm}%
\usepackage{mathrsfs}%
\usepackage[title]{appendix}%
\usepackage{xcolor}%
\usepackage{textcomp}%
\usepackage{manyfoot}%
\usepackage{booktabs}%
\usepackage{algorithm}%
\usepackage{algorithmicx}%
\usepackage{algpseudocode}%
\usepackage{listings}%
\usepackage{nameref}%
\usepackage[justification=justified]{caption}
\usepackage{chngcntr} 
\usepackage[capitalise]{cleveref}%

\makeatletter
\newcommand\setcurrentname[1]{\def\@currentlabelname{#1}}
\makeatother

\input{00_definitions}

\begin{abstract}
    Science is a collaborative endeavor in which ``who collaborates with whom'' profoundly influences scientists' career trajectories and success.
    Despite its relevance, little is known about how scholars facilitate new collaborations among their peers.
    In this study, we quantify brokerage in academia and study its effect on the careers of physicists worldwide.
    We find that early-career participation in brokerage increases later-stage involvement for all researchers, with increasing participation rates and greater career impact among more successful scientists.
    This cumulative advantage process suggests that brokerage contributes to the unequal distribution of success in academia.
    Surprisingly, this affects both women and men equally, despite women being more junior in all brokerage roles and lagging behind men's participation due to their late and slow arrival to physics. Because of its cumulative nature, promoting brokerage opportunities to early career scientists might help reduce the inequalities in academic success.
\end{abstract}

\begin{document}

\flushbottom

\maketitle

\input{01_introduction}

\input{02_results}
\input{03_discussion}
\input{04_methods.tex}

\input{05_data_code_availability}

\bibliography{Motifs}

\input{06_acknowledgements}

\appendix

\clearpage
\setcounter{figure}{0}
\renewcommand{\contentsname}{Supplementary Notes}
\renewcommand{\listfigurename}{Supplementary Figures}
\renewcommand{\thefigure}{S\arabic{figure}}
\renewcommand{\thesection}{S\arabic{section}}
\crefname{appendix}{Section}{Sections}
\Crefname{appendix}{Section}{Sections}
\crefname{subappendix}{Section}{Sections}
\Crefname{subappendix}{Section}{Sections}

\tableofcontents
\listoffigures
\clearpage

\input{si_00_dataset}
\clearpage
\input{si_01_intermediate_brokerage}
\clearpage
\input{si_02_careers}
\clearpage
\input{si_03_metrics}
\clearpage
\input{si_04_gender_impact}
\clearpage
\input{si_05_normalization}

\end{document}

%% file: 00_definitions.tex
\newcommand{\Nauthd}{430,755} 
\newcommand{\Nauthf}{138,124} 
\newcommand{\Npapers}{617,504} 
\newcommand{\Nauthorships}{1,941,574} 
\newcommand{\ThreshGender}{66\%} 

\newcommand{\Npermu}{5,000} 

\newcommand{\Nauthunid}{200,669} 
\newcommand{\Nauthid}{230,086} 
\newcommand{\Nauthidwoman}{34,596} 
\newcommand{\Pauthidwoman}{15\%} 

\title{Cumulative advantage of brokerage in academia}

\author[1,2,$*$]{Jan Bachmann}
\author[1,2]{Lisette Esp\'{i}n-Noboa}

\author[3,2,4]{Gerardo I\~niguez}

\author[5,1,$*$]{Fariba Karimi}

\affil[1]{Complexity Science Hub Vienna, 1080 Vienna, Austria}
\affil[2]{Department of Network and Data Science, Central European University, 1100 Vienna, Austria}

\affil[3]{Faculty of Information Technology and Communication Sciences, Tampere University, 33720 Tampere, Finland}
\affil[4]{Centro de Ciencias de la Complejidad, Universidad Nacional Autonóma de México, 04510 Ciudad de México, Mexico}

\affil[5]{Graz University of Technology, 8010 Graz, Austria}

\affil[$*$]{\small{\textit{Corresponding authors email: bachmann@csh.ac.at, karimi@tugraz.at}}}

%% file: 01_introduction.tex
\section*{Introduction}
\label{sec:intro}
Academic achievements often emerge from collaborative efforts that combine diverse perspectives,~\cite{sun.etal_socialdynamicsscience_2013} foster the exchange of ideas, delegate tasks, and create networks for disseminating information.~\cite{granovetter_strengthweakties_1973,burt_structuralholessocial_1995}
Collaborative teams may form by repeated~\cite{petersen_quantifyingimpactweak_2015} or new collaborations,~\cite{guimera.etal_teamassemblymechanisms_2005} each offering distinct benefits such as increased productivity and more innovation.
When looking for new collaborative bonds, scholars often rely on their existing academic network through referrals by trusted collaborators.~\cite{newman_structurescientificcollaboration_2001,beaver_reflectionsscientificcollaboration_2001}
An introduction where a trusted intermediary introduces two former collaborators is known as \textit{brokerage}.~\cite{obstfeld_socialnetworkstertius_2005}

As collaborative teams grow in relevance,~\cite{wuchty.etal_increasingdominanceteams_2007} it is not only the number of co-authors that matters but also who those collaborators are.
For instance, jointly publishing with senior researchers in high-impact journals significantly boosts junior researchers' chances of publishing in the same journals later on in their careers.~\cite{sekara.etal_chaperoneeffectscientific_2018}
The centrality of scientists in collaboration networks~\cite{newman_structurescientificcollaboration_2001} predicts their future impact, as measured by subsequent citations received.~\cite{sarigol.etal_predictingscientificsuccess_2014}
Highly productive collaborations, known as super-ties, can further boost impact.~\cite{petersen_quantifyingimpactweak_2015}
Both productivity and the amount of highly cited papers depend on a scientist's collaborators.~\cite{li.etal_untanglingnetworkeffects_2022}
And yet, brokerage as a process through which collaborations are formed, and especially its relation to academic achievements, remains largely unexplored.

Brokerage is crucial for innovation,~\cite{obstfeld_socialnetworkstertius_2005,lingo.omahony_nexusworkbrokerage_2010} bridging between the resourceful core and the innovative periphery of collaborative networks.~\cite{juhasz.etal_brokeringcoreperiphery_2020}
In the public sector, brokerage can create opportunities for even more brokerage afterwards.~\cite{kauppila.etal_openingnewbrokerage_2024}
This suggests that brokerage in academia might be a cumulative process, akin to the Matthew effect often observed in science,~\cite{merton_mattheweffectscience_1968} where highly cited papers receive even more citations in the future.
Hence, brokerage might have long-lasting implications on scientists' careers, contributing to the concentration of success among a few individuals while leaving the majority of scholars less recognized.~\cite{redner_howpopularyour_1998,merton_mattheweffectscience_1968,petersen.etal_quantitativeempiricaldemonstration_2011}

As a time-sensitive process~\cite{obstfeld.etal_brokerageprocessdecoupling_2014}, brokerage likely disadvantages women, who are still under-represented in senior positions.
Their late-entry into academia~\cite{kong.etal_influencefirstmoveradvantage_2022} and higher dropout rates~\cite{spoon.etal_genderretentionpatterns_2023} partly contribute to the persistent gender disparities in publication numbers~\cite{huang.etal_historicalcomparisongender_2020} and received citations.~\cite{kong.etal_influencefirstmoveradvantage_2022}
Although a global phenomenon, the gender gap is particularly strong in STEM-related fields, such as physics.~\cite{huang.etal_historicalcomparisongender_2020,lariviere.etal_bibliometricsglobalgender_2013}
The choice of collaborators is another contributing factor to these inequalities.~\cite{li.etal_untanglingnetworkeffects_2022}
Women tend to collaborate with other women,~\cite{karimi.oliveira_inadequacynominalassortativity_2023,sajjadi.etal_unveilinghomophilypool_2024,jadidi.etal_genderdisparitiesscience_2017} forming clustered communities,~\cite{jadidi.etal_genderdisparitiesscience_2017} which might structurally limit their access to new collaborations.
Participating in brokerage in these clusters may worsen disparities by pushing women to the periphery and exacerbating their segregation.~\cite{asikainen.etal_cumulativeeffectstriadic_2020}
Therefore, a deeper understanding of the interplay between gender and brokerage in academia requires further investigation.

\begin{figure}[t]
    \centering
    \includegraphics{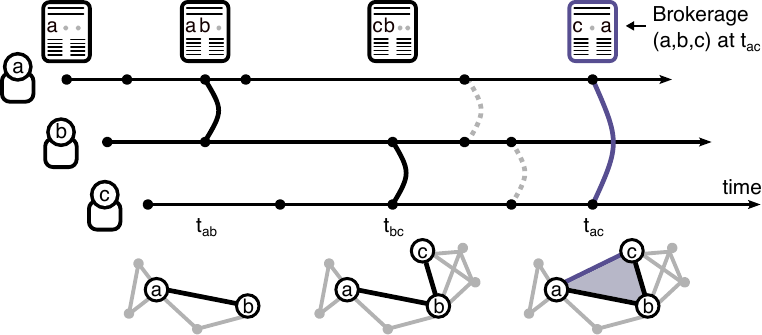}
    \caption[]{\small\textbf{Brokerage in academic collaborations}.
        Joint publications among three authors $a$, $b$ and $c$ create links in the collaboration network at the time of their first publication (solid, curved arcs with an aggregated view at the bottom).
        We consider the collaboration between $a$ and $c$ at time $t_{ac}$, with or without $b$, as the {\it tertius iungens}~\cite{obstfeld_socialnetworkstertius_2005} brokerage event between $a$, $b$ and $c$.
        At this point, the broker $b$ has collaborated separately with $a$ at $t_{ab}$ and $c$ at $t_{bc}$ before this joint publication.
        Repeated collaborations (dashed gray arcs) are disregarded in determining the brokerage event between the three co-authors.}
    \label{fig:01-01_fig1}
\end{figure}

To this end, we quantify the impact of scientists' brokerage participation on their eventual accumulated achievements.
In a large-scale temporal network, constructed from collaborations in the field of physics, we define brokerage as the initiation of links between two authors who have not yet collaborated but share a common former collaborator, the broker (\cref{fig:01-01_fig1}).
We investigate the relationship between brokerage activity throughout a scientist's career and their eventual scientific achievements, as measured by their total number of publications and received citations within the American Physical Society (APS) journals.
We then verify whether brokerage is cumulative by comparing individual brokerage participation across five career stages, and conclude by analyzing women's representation in brokerage.

Our results show that the more a scientist has achieved by the end of their career, the more frequently they have participated in brokerage before.
Only the most successful scientists accelerate the rate at which they participate in brokerage.
Engaging in brokerage during junior career stages tends to lead to more brokerage involvement at senior stages, creating a cumulative effect that is particularly strong for the most successful scientists.
Surprisingly, these effects apply equally to men and women, even though women's late and slow arrival is reflected in their brokerage participation.
Women tend to be more junior in all brokerage roles, and the first occurrence of all-women brokerage lags behind the first female publication by 80 years, which is 50 years longer than the delay between the first male publication and all-men brokerage.

\begin{figure}[ht]
    \centering
    \includegraphics{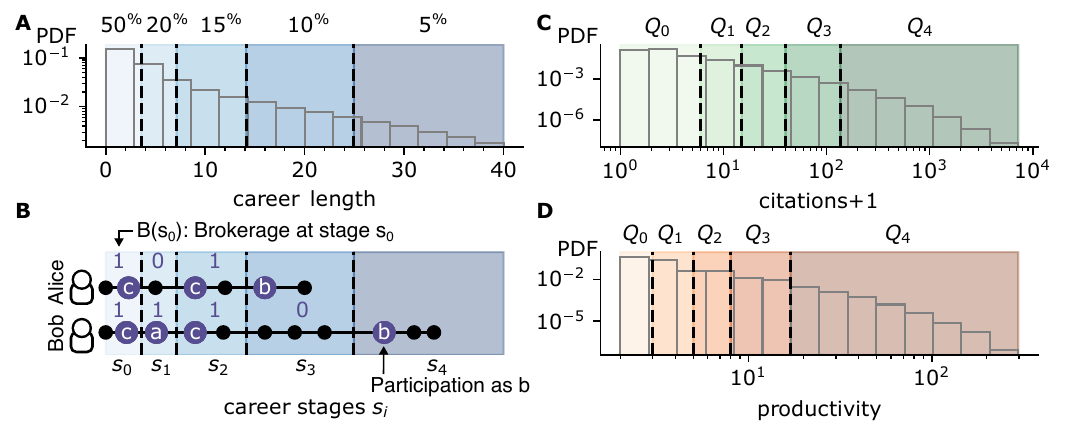}
    \caption[]{\small\textbf{Skewed career lengths and academic impact}.
    (\textbf{A}) The distribution of career lengths, as measured by the years between the first and last publication. To account for variations in career lengths, we partition this distribution into five percentile-based bins of decreasing size which we refer to as career stages $s_0$ to $s_4$ (color shades, sizes on top).
    (\textbf{B}) Throughout these stages, scientists publish papers and may participate in brokerage multiple times in varying roles (as $a$, $b$, or $c$; see~\cref{fig:01-01_fig1}). We count their brokerage participation at each completed stage $s_i$ as $B(s_i)$.
    For example, both Alice and Bob participated once in their first career stage, $B(s_0) = 1$.
    (\textbf{C--D}) The distribution of academic impact by the end of a career is measured by the total number of received citations and publications.
    To compare scientists by their career impact, we place them in an impact group $Q_0$ to $Q_4$.
    As with the career stages, we partition the impact distributions using the same five percentile-based bins (color shades).
    For example, Alice achieved $Q_2$ in productivity (D) with a total of seven articles (circles in her timeline in B), while Bob achieved $Q_3$ with eleven publications.
    Overall, the vast majority of scientists leave academia early and achieve a lower impact in their scientific careers.
    A small fraction, however, stay longer and achieve significant success.
    Citation counts were increased by 1 only for the logarithmic scale visualization without affecting the actual percentile binning.
    }
    \label{fig:01-02_metrics_hist}
\end{figure}

%% file: 02_results.tex
\section*{Results}
\label{sec:results}

\begin{figure}[t]
    \centering
    \includegraphics{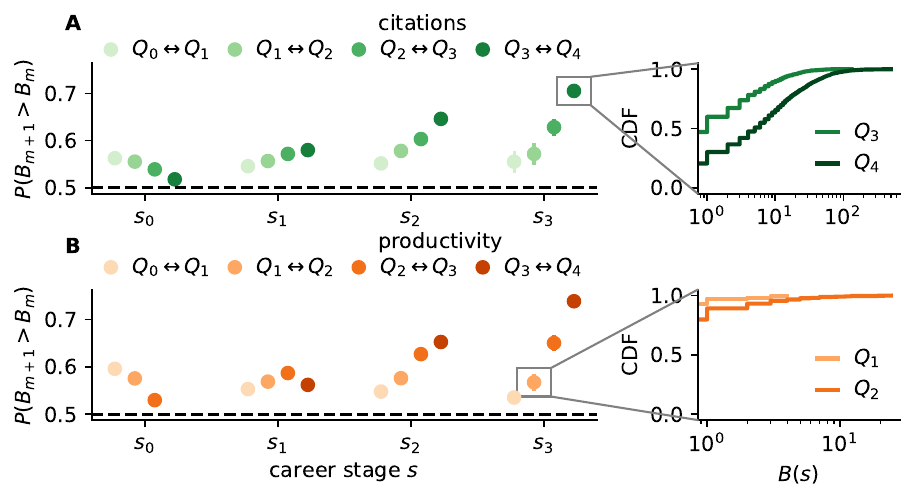}
    \caption[]{\small\textbf{Brokerage frequency and academic impact}.
    We measure academic impact separately for two metrics: (\textbf{A}) the total number of received citations and (\textbf{B}) the total number of publications.
    We compare the brokerage participation $B_m(s_i)$ of scientists in impact group $Q_m$ to those that achieve the next impact level $Q_{m+1}$ (increasing color intensity) along career stages $s_i$ (x-axis).
    The cumulative distributions of brokerage for two different impact groups at stage $s_3$ are shown on the right.
    Each marker on the left shows the probability $P(B_{m+1} > B_m)$ that a scientist from group $Q_{m+1}$ has a higher brokerage frequency than a scientist from group $Q_m$.
    Error bars indicate bootstrapped $95\%$-confidence intervals, and all visible results are statistically significant ($p<0.05$, see \nameref{sec:methods}).
    Almost all comparisons are above the neutral line, showing a positive relation between participation in brokerage and eventual success across all career stages.
    Scientists who eventually reach higher success levels engage in brokerage more frequently across all career stages.
    While this effect remains positive but roughly constant for the least successful scientists, it increases for the most successful ones over their careers.
    Comparing the two most successful groups ($Q_3 \leftrightarrow Q_4$), small differences in early careers ($s_0$) snowball later on ($s_3$), hinting at a cumulative nature of brokerage.
    }
    \label{fig:03_bf_success}
\end{figure}

\subsection*{Measures of brokerage and academic impact}
\label{ssec:brokerage_process}
Following the definition by Obstfeld, we refer to {\it tertius iungens} brokerage (brokerage, for short) as ``a strategic behavioral orientation toward connecting people in their social network by either introducing disconnected individuals or facilitating new coordination between connected individuals”.~\cite{obstfeld_socialnetworkstertius_2005}
Therefore, we track brokerage events as joint publications of two scientists $a$ and $c$, who have each previously and independently collaborated with a third author $b$, the broker (\cref{fig:01-01_fig1}).
We analyze brokerage in the context of individual careers, focusing on completed careers having a minimum of two publications and lasting less than 40 years (see \nameref{sec:methods} for more details).
We identify \Nauthf~authors who participate in brokerage an average of $1.3$ times throughout their careers.
To account for the heterogeneity in career lengths, we partition careers into five stages ($s_0$ to $s_4$) based on a percentile-based binning of the career length distribution (\cref{fig:01-02_metrics_hist}A--B and \nameref{sec:methods}).
The binning reveals the strong heterogeneity of career lengths: half of the population stops publishing after less than four years, while only $5\%$ of physicists have careers lasting more than 25 years (see first and last shaded areas in~\cref{fig:01-02_metrics_hist}A).
Within each stage $s_i$, we count the number of brokerage events $B(s_i)$ in which a scientist has participated (\cref{fig:01-02_metrics_hist}B).
We then quantify the academic impact at the end of a career by assigning an impact group based on the percentile bin achieved.
This classification considers either the number of received citations or the total number of publications, categorized from $Q_0$ (least successful) to $Q_4$ (most successful) (\cref{fig:01-02_metrics_hist}.C--D).

\subsection*{The role of brokerage in academic impact}
We collect the distributions of brokerage frequencies $B_m(s_i)$ at a given stage $s_i$ for all impact groups $Q_m$.
Considering the broadness of these distributions (see Supplementary Information [SI] Fig. S5), we calculate the probability $P(B_{m+1}(s_i) > B_m(s_i))$ that an individual's brokerage frequency in the more successful group $Q_{m+1}$ exceeds that of a randomly chosen individual in the less successful group $Q_m$ (see SI Fig. S7 for an alternative metric).
This allows us to determine if more successful scientists tend to participate more in brokerage or not.

Across all impact groups, scientists in the more successful group, $Q_{m+1}$, tend to engage in more brokerage than their less successful counterparts, $Q_m$. This is evidenced by probability values exceeding $0.5$, the neutrality point (see left panels in \cref{fig:03_bf_success}A-B).
Between the least successful scientists, $Q_0 \leftrightarrow Q_1$, the difference in brokerage participation remains roughly constant across career stages and impact metrics, averaging around $0.55$ for both citations and productivity (see the lightest data points in \cref{fig:03_bf_success}A-B).
The trend changes at higher impact levels, particularly when comparing the two most successful groups, $Q_3$ and $Q_4$.
In these groups, the differences in brokerage amplify, ranging from approximately $0.5$ to $0.7$ for both citations and productivity (see the darkest data points in \cref{fig:03_bf_success}A-B), indicating that small differences in early career stages can determine brokerage opportunities in later stages.

Within the earliest career stage, the differences in brokerage are more pronounced among the least successful groups.
Specifically, the difference in brokerage participation between $Q_1$ and $Q_0$ is the largest compared to any other pair of groups (see negative trend at $s_0$ in~\cref{fig:03_bf_success}A-B).
While this difference remains constant in later career stages, the differences between the more successful groups increase considerably, especially in $s_2$ and $s_3$.
For instance, in the most senior career stage, the difference in brokerage participation between $Q_4$ and $Q_3$ is the largest compared to any less successful pair (see positive trend at $s_3$ in \cref{fig:03_bf_success}A--B).
These results account for group size differences across stages and impact groups, and hold across cohorts starting in different decades (SI Fig. S9).
Scientists who eventually become more successful participate in brokerage more often throughout their careers.
Taken together, brokerage has a clear positive effect on academic impact in all career stages, and this effect intensifies with increasing success levels and with later career stages.

\begin{figure}[t]
    \centering
    \includegraphics{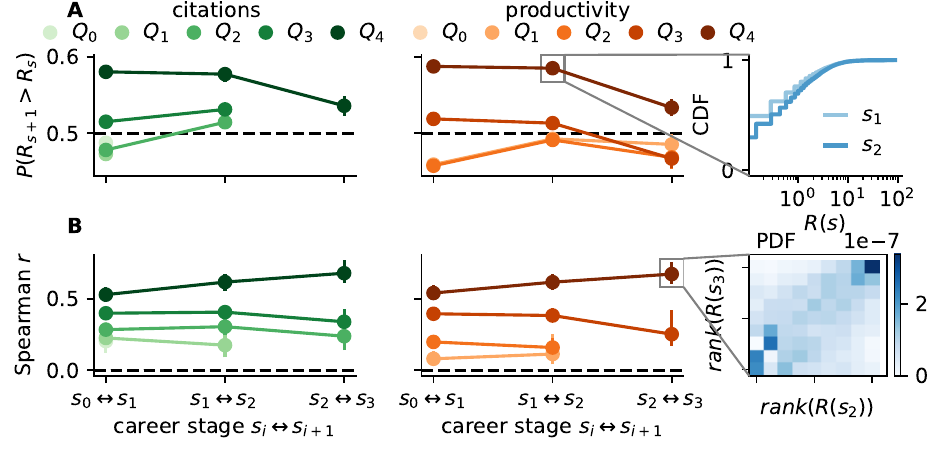}
    \caption[]{\small\textbf{Cumulative advantage in brokerage}.
    (\textbf{A})
    Brokerage rate changes across consecutive career stages.
    Each marker shows the probability $P(R_{s+1} > R_s)$ that a scientist of impact group $Q_m$ has a higher brokerage count per year at stage $s+1$ compared to the earlier stage $s$.
    Statistical test and uncertainty estimations are the same as in~\cref{fig:03_bf_success}.
    We find that the rates increase with each career stage for the most successful scientists, indicating that they not only accumulate more brokerage but also do so at increasing rates.
    A consistent speed-up is also unique to them, given that less successful groups show a significant slow-down of brokerage throughout many career stages, especially in productivity.
    The cumulative distributions of brokerage rates for $Q_4$ at stage $s_1$ and $s_2$ are shown on the right.
    (\textbf{B}) The cumulative nature of brokerage rates, as measured by a positive correlation of individuals' rates between two consecutive career stages $s_i$ and $s_{i+1}$.
    Scientists with a higher brokerage rate in one stage tend to have a higher rate in the next stage and vice versa.
    This correlation is also stronger for more successful researchers, indicating that the cumulative nature of brokerage is more pronounced for them.
    The rank correlation between brokerage rates at stages $s_2$ and $s_3$ of scientists in productivity $Q_4$ is shown by the joint PDF on the right.
    }
    \label{fig:04_bfr}
\end{figure}

\subsection*{Speed-up and slow-down of brokerage participation}
As scholars progress through career stages, their brokerage activity may change.
To compare brokerage participation across stages of varying duration, we compute the annual brokerage rate as the fraction of the brokerage frequency and the duration of each career stage (see \nameref{sec:methods}).
Then, for each impact group $Q_m$, we compare the brokerage rates of scientists in that group across two consecutive career stages $s_i \leftrightarrow s_{i+1}$ (\cref{fig:04_bfr}A).
A constant rate would lie on the neutral line, while a positive value indicates an increase in brokerage rates over the two consecutive career stages (see SI Fig. S8A for an alternative metric).
Using the same statistical framework as before, we find that most scientists show identical rates or even a slowdown in brokerage rates over career stages.
This is striking given that, on average, the rate of productivity, as measured by annual publications, remains constant or increases throughout a career.~\cite{sinatra.etal_quantifyingevolutionindividual_2016}
The brokerage slow-down diminishes over growing career success, transitioning fully to a speed-up only for the most successful scientists (\cref{fig:04_bfr}A, comparing $Q_4$ to other impact groups).
In contrast to the rest of this physics community, $Q_4$ consistently increases their brokerage rate, suggesting that brokerage could be a decisive factor in reaching the very top of the academic impact distribution.

So far, we have shown that brokerage is beneficial for academic impact and that the most successful scientists consistently increase their brokerage rates over time.
Both findings suggest that brokerage is a cumulative process, where brokering collaborations at one point creates more brokerage opportunities later on.
To test whether brokerage is indeed cumulative for individual scientists, we compute the correlation of their rates between two consecutive career stages $s_i$ and $s_{i+1}$.
We compute the Spearman correlation coefficient of these rates for each success group (\cref{fig:04_bfr}B).
The results are consistent when using the Pearson correlation coefficient (SI Fig. S8B).
The positive correlation indicates that brokerage in general is cumulative: those who have a higher brokerage rate among comparably successful scientists (same impact group) in one stage tend to have a higher rate in the next stage (and vice versa).
Our results support previous conjectures on the cumulative nature of brokerage identified in governmental and tech industry organizations.~\cite{kauppila.etal_openingnewbrokerage_2024}
The correlation further increases with eventual impact.
More successful scientists appear to be better at capitalizing on brokerage opportunities, which leads to a more pronounced cumulative advantage in brokerage for them.

\begin{figure}[t]
    \centering
    \includegraphics[width=\textwidth]{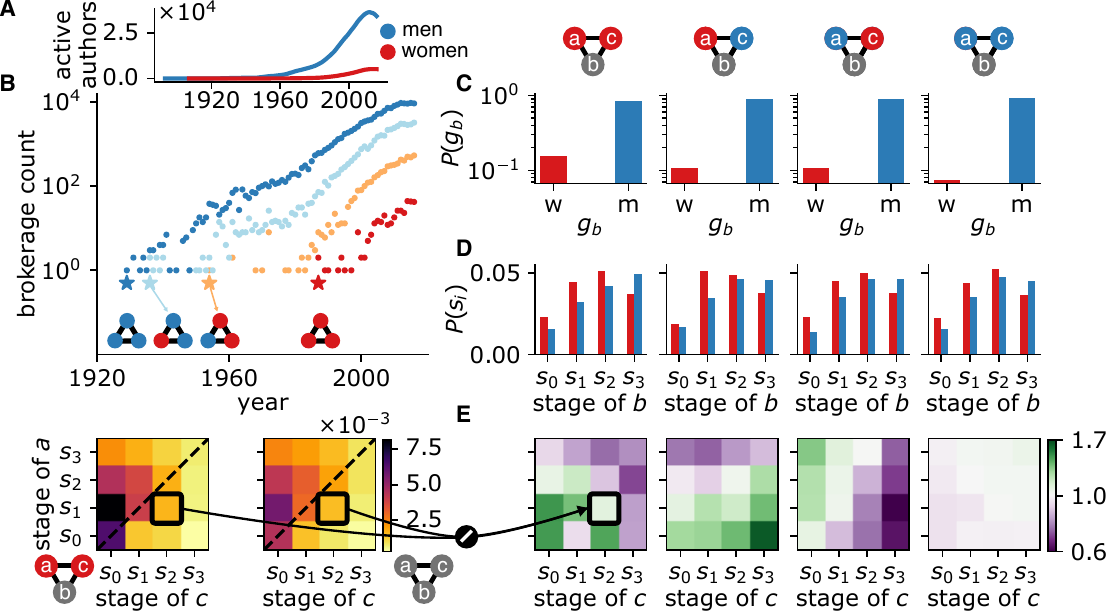}
    \caption[]{\small\textbf{Gender disparities in brokerage participation}.
    (\textbf{A}) The number of active female and male authors per year shows the under-representation of women in physics.
    (\textbf{B}) Blue dots represent all-men brokerage event counts, and red dots represent all-women events. Light blue dots indicate events with two men and one woman, and light red dots indicate events with two women and one man. 
    Stars mark the first-ever occasion of a given type of gendered brokerage.
    We see that all-women brokerage started appearing roughly 80 years after the first women joined physics.
    (\textbf{C}) The probability of a broker being a woman (w) or a man (m) given the gender of co-authors $a$ and $c$ (columns).
    Men are much more likely to be brokers. Women are less likely to be brokers if $a$ or $c$ are men.
    (\textbf{D})
    Women usually serve as brokers earlier in their careers, while men typically do so at more senior levels.
    (\textbf{E}) Comparison of joint career stage distributions of authors $c$ (x-axis) and $a$ (y-axis).
    We divide the distribution of a given gender composition (columns) by the distribution over all brokerage events, regardless of gender (the case where both are female is shown in the left inset).
    Values over 1 (green) indicate an over-representation of the respective gender-seniority combination.
    Values below 1 (purple) mean under-representation.
    We find that the differences in representation depend on the roles in which women participate.
    Whenever $a$ or $c$ are female, there is an under-representation in the respective senior career stages (rows for $a$ and columns for $c$ in the three left-most panels).
    When both $c$ and $a$ are male, their distribution is similar to the overall distribution, as they contribute the highest group size (the right-most panel in E).
    }
    \label{fig:05_gender}
\end{figure}

\subsection*{Gender disparities in brokerage}
\setcurrentname{Gender disparities}
\label{ssec:gender_disaprities}
To investigate whether the long-term effects of brokerage differ by gender, we build upon our previous findings---which showed that the impact of brokerage accumulates differently across career stages---and consider the late entry of women into physics (\cref{fig:05_gender}A).
We find that women's late arrival in the field strongly determines the occurrence of brokerage events.
For instance, although women's first APS publication was in 1907 (fourteen years after men's first publication), it took 47 years for the first brokerage to predominantly include women and 80 years for the emergence of all-women brokerage events (red stars in~\cref{fig:05_gender}B).
When normalizing these brokerage counts by the group sizes of active women and men, we find that the disparities are most likely due to the low representation of women, rather than a behavioral difference in women's engagement in brokerage (SI Note S6 and Fig. S13).
Similarly, when comparing impact groups per gender and controlling for varying career lengths (a major driver of gender disparities on academic impact),~\cite{huang.etal_historicalcomparisongender_2020} we find that the benefits of brokerage for academic impact are similar for both women and men (SI Figs. S10--S12).

The dependency of brokerage engagement on gender group size motivates us to investigate other representational disparities in the seniority and roles of the participating authors.
Given the genders of authors $a$ and $c$, we find a predominance of male brokers $b$ (\cref{fig:05_gender}C) in all gender combinations.
However, when comparing the proportion of female brokers, the all-women brokerage type shows the highest percentage of female brokers, while the composition with both $a$ and $c$ being male shows the least.
This could be an effect of the gender homophily, a preference for same-gendered collaborations, observed in the field of physics.~\cite{jadidi.etal_genderdisparitiesscience_2017,karimi.oliveira_inadequacynominalassortativity_2023,sajjadi.etal_unveilinghomophilypool_2024}
The seniority distribution of brokers by gender (\cref{fig:05_gender}D) shows that men tend to be in late career stages, while women are typically in early- to mid-career stages.
To better understand the role of authors $a$ and $c$, we investigate the interplay between their gender and seniority (\cref{fig:05_gender}E).
By combining the different career stages of $a$ and $c$, we identify 16 career stage pairs and calculate their joint distribution.
We then assess their relative frequency by computing the ratio of the distribution of a specific gender combination and the overall distribution regardless of gender (see the heatmaps in \cref{fig:05_gender}E's inset, and~\nameref{sec:methods} for details).
The seniority representation mainly depends on the participation of women.
When two women engage in brokerage as $a$ and $c$, they are less likely to be in senior career stages (last row and column in the left-most panel of \cref{fig:05_gender}E).
If women and men participate together, we observe an under-representation in senior career stages for the role of the female scientist (rows in the second panel and columns in the third panel of \cref{fig:05_gender}E).
Brokerage events in which a senior male scientist in role $c$ is introduced to a junior female scientist in role $a$ are over-represented the most, while the case in which the genders are swapped is the least represented one (dark green and purple cells in the second and third panel of \cref{fig:05_gender}E).
When both are men (the right-most panel in \cref{fig:05_gender}E), the distribution closely aligns with the general distribution (values close to one), since predominantly male brokerage events constitute the majority of all brokerage counts.
These results suggest that the age at which scientists participate in brokerage, regardless of their role as $a$, $b$, or $c$, can be anticipated by their gender.
This is likely due to the late entry of women into physics and academia, and their high drop-out rates in senior career stages.~\cite{huang.etal_historicalcomparisongender_2020,spoon.etal_genderretentionpatterns_2023}

%% file: 03_discussion.tex
\section*{Discussion}
\label{sec:discussion}
Scientific collaboration plays a fundamental role in academia and how recognition is distributed.
A publication's future impact can be predicted from the author's position in their collaboration network~\cite{sarigol.etal_predictingscientificsuccess_2014} as beneficial research practices are transferred through collaborative ties from senior to junior researchers.~\cite{li.etal_untanglingnetworkeffects_2022}
Previous studies show that new collaborations~\cite{guimera.etal_teamassemblymechanisms_2005} balance the benefits and trade-offs of repeated collaborations.~\cite{petersen_quantifyingimpactweak_2015}
In this study, we quantify brokerage~\cite{obstfeld_socialnetworkstertius_2005} as the initiation of new collaborations through shared former co-authors and measured its influence on academic impact.

By analyzing a collaboration network spanning over a century of publications in the field of physics, we identified more than 130k scientific careers, revealing diverse brokerage patterns across different career stages and genders.
In particular, we found a positive correlation between participation in brokerage throughout all stages of a scientific career and the academic impact achieved at its culmination.
The strength of this effect varies with the career stage and the final level of impact achieved.
When comparing the least successful scientists, the difference in brokerage participation between them remains constant over time.
In contrast, for more successful scientists, small differences early on turn into a strong increase over their careers.
Only researchers with the most impact manage to accelerate their participation in brokerage, while less productive scholars experience a slowdown instead.
We find that a mechanism of cumulative advantage can explain how small differences early on lead to an increasing gap in outcomes over time, implying that early brokerage stimulates more brokerage in the future.
Women make up only \Pauthidwoman~of authors and are typically at more junior stages in their careers when they participate in brokerage.
Despite this, women receive the same academic impact benefits from brokerage as men.
Additionally, all-female brokerage occurs later compared to brokerage involving other gender compositions.
Both of these findings are most likely caused by the under-representation of women both in academia~\cite{kong.etal_influencefirstmoveradvantage_2022} and in senior roles due to higher dropout rates.~\cite{spoon.etal_genderretentionpatterns_2023}

We utilize large-scale, longitudinal data spanning entire careers to provide observational insights, thereby contributing to the existing literature on brokerage in creative work with a specific focus on academia.
Creative work often necessitates collaboration,~\cite{hargadon.bechky_whencollectionscreatives_2006} and brokerage plays a crucial role in facilitating this process.~\cite{lingo.omahony_nexusworkbrokerage_2010,obstfeld_socialnetworkstertius_2005}
For example, music producers act as brokers by integrating the contributions of others, a crucial step for the successful execution of the innovation process.~\cite{lingo.omahony_nexusworkbrokerage_2010}
Brokers gain additional benefits by bridging the gap between film directors at the core and periphery of creative collaboration networks in movie co-creation. This positioning potentially facilitates access to resources and increases visibility in the core, while also bringing innovative ideas from the periphery.~\cite{juhasz.etal_brokeringcoreperiphery_2020}
In academia, brokerage is likely to be beneficial for the same reasons.
Facilitating new collaborations among scholars may lead to the formation of new ideas and discoveries by leveraging diverse expertise.~\cite{sun.etal_socialdynamicsscience_2013}
The successful execution of scientific endeavors requires resources that tend to concentrate on already well-established scholars.~\cite{merton_mattheweffectscience_1968}
By introducing senior scientists to new collaborators, brokerage may increase the visibility of junior scientists and ensure access to new perspectives and innovative ideas for more experienced scholars.
A future study could investigate the interplay between all participants' success, seniority and network position to further understand the mechanisms that make brokerage beneficial in academia.

Brokerage may provide an additional benefit to the participating scientists by bringing their respective collaborators closer together.
This creates the opportunity for subsequent brokerage.
By introducing their new collaborators to their former colleagues, or the former broker, the introduced scientists may become brokers themselves.
This leads to a consolidation of the local neighborhood which is a typical property of scientific collaboration networks.~\cite{newman_structurescientificcollaboration_2001}
Analyzing such chains of brokerage could reveal why brokerage is cumulative, thereby complementing our findings.
Combined with the benefits it brings, brokerage is thus likely to contribute to the processes of cumulative advantage that drive the unequal distribution of academic success.~\cite{merton_mattheweffectscience_1968}
As these processes exacerbate small differences over time, promoting brokerage opportunities to early career scientists might reduce the inequalities in academic impact that unfold over scientific careers.

By limiting our study to co-authorships and their temporal order within APS journals, we are potentially ignoring other channels of introduction such as meetings at conferences, research visits, or other journals.
Scientists might introduce each other without ever publishing together, or, if they do collaborate, their article might be published at a later time.
While we account for many confounders in assessing brokerage's effect on career success, other factors like topic popularity, community size, and publication ease may also play a role.
On a macroscopic scale, the rapid growth of publication counts~\cite{fire.guestrin_overoptimizationacademicpublishing_2019} and the increasing dominance of collaboration teams~\cite{wuchty.etal_increasingdominanceteams_2007} could further influence the way scientists engage in brokerage.
While our results are robust throughout various decades, future work should focus on adding a multi-layer perspective to include formal and informal relationships, co-evolving processes and their impact on brokerage and success.

While participation in brokerage is mostly beneficial, the strength of its benefit and its rate throughout a career depends on how successful someone will eventually become.
This suggests that brokerage is an essential part of a successful academic career.
A comprehensive understanding of the role of brokerage can motivate scientists to actively engage in it, and deepen our insights into how inequality in academic success emerges and persists.
This paper lays the groundwork for future research on mitigating these inequalities by fostering brokerage opportunities for all scientists, particularly those in the early stages of their careers.

%% file: 04_methods.tex
\section*{Methods}
\setcurrentname{Methods} 
\label{sec:methods}
\hypertarget{ssec:data_processing}{\subsection*{Data description}}
We consider $\Npapers$ articles published in journals of the American Physical Society (APS) between 1893 and 2021.
Building on an existing method,~\cite{sinatra.etal_quantifyingevolutionindividual_2016} we use authors' names, affiliations, and citation practices, to address the so-called name disambiguation problem, tracking papers over an author's career and distinguishing authors with identical names.
To address first name variations, we ensure consistency across an author's papers; otherwise, we treat them as different authors (see SI Note S1 and Fig. S1--S2 for details).
For example, we would treat authors L. Smith, Laura Smith, and Leonard Smith as three distinct authors, as L. could stand for either Laura or Leonard.
This additional step might come at the cost of falsely removing authorship (i.e., paper-author links) from scientists whose first names could not be disambiguated.
However, it provides a more realistic count of author names used per author and more accurate results (SI Fig. S3 and Note S1).
In addition, this error results in a conservative upper bound for the brokerage counts, since we will overlook real brokerage events as opposed to considering events that never happened.

This process yields a total of \Nauthd~disambiguated authors and \Nauthorships~authorships, from which we construct a temporally growing collaboration network.
In this graph, authors are represented as nodes, and links are created between them at the time they publish a co-authored paper (see~\cref{fig:01-01_fig1}).
We order authorships by the publication date and refer to the timestamp of a new link between two authors, $x$ and $y$, created through such a publication as $t_{xy}$.

\subsection*{Career length and career stages}
Given the heterogeneity of career lengths, measured by the years between the first and last publication (\cref{fig:01-02_metrics_hist}A), we partition scientific careers into five stages, $s_i$.
As border values of the stages, we use the $50^{th}$, $70^{th}$, $85^{th}$, and $95^{th}$ percentiles of the career length distribution (\cref{fig:01-02_metrics_hist}B).
Stage $s_0$ represents the first 50\% of scientists with the shortest career lengths of less than 3.6 years.
Stage $s_1$ refers to the next 20\% with career lengths between 3.6 and 7.1 years, followed by $s_2$, the next 15\% of scientists whose careers span between 7.1 and 14.1 years.
Career stage $s_3$ corresponds to the next 10\% of scientists who published between 14.1 and 24.9 years, and $s_4$ the last 5\% of researchers who kept publishing for at least 24.9 years.
The respective duration in years for each career stage $s_0$ to $s_4$ are:
$\Delta t_{s_0}=3.6$,
$\Delta t_{s_1}=3.5$,
$\Delta t_{s_2}=7$,
$\Delta t_{s_3}=10.8$,
$\Delta t_{s_4}=15.1$.
This binning accounts for the heterogeneity in career lengths (\cref{fig:01-02_metrics_hist}A) and it addresses the correlation between career length and cumulative brokerage counts (SI Fig. S4) by which those with longer careers accumulate more brokerage.
The inferred stages align with the binning method used in a comparable study.~\cite{milojevic.etal_changingdemographicsscientific_2018}

To illustrate the career stages defined above, consider the example in~\cref{fig:01-02_metrics_hist}B.
We observe that Alice reached career stage $s_3$ and Bob reached career stage $s_4$.
In our analysis, when gathering statistics at career stage $s_2$, we include the brokerage activities of both Alice and Bob at this particular stage.
However, when retrieving statistics in $s_3$ or $s_4$, Alice is excluded as her career ended at stage $s_3$.

We exclude certain careers defined as outliers as follows.
We remove authors with only one publication, as our analysis requires at least two publications to calculate the career length.
Very long careers, those who were active for over 40 years, were also excluded, as these are likely errors from merging different scientists with very similar names.
We also exclude scientists who are still active at the end of the observation period, keeping only those who stopped publishing by the end of 2016.~\cite{sinatra.etal_quantifyingevolutionindividual_2016}
Including active authors, who have potentially not finished their careers, would underestimate their final impact and assign them to low-impact categories.
Our results are robust to this threshold because they hold for authors who started their careers more than 40 years before 2016 (see SI Fig. S9).
Out of all disambiguated authors, the combined filtering steps leave a total of $\Nauthf$ careers, summarized in~\cref{fig:01-02_metrics_hist}.

\subsection*{Measuring academic impact}
We quantify the academic impact of each scientist by the end of their career by using two metrics: the total number of publications and the total number of received citations.~\cite{fortunato.etal_sciencescience_2018}
We only consider publications and citations within APS journals.
Instead of using raw counts in our analysis, we assign five impact categories, $Q_0$ to $Q_4$, to each author based on a percentile-based binning.
Similar to the career stages, we partition both the citations and publications distributions by their $50^{th}$, $70^{th}$, $85^{th}$, and $95^{th}$ percentiles.
These percentiles serve as categories to determine the achieved career impact for each author.
The respective impact values of the groups $Q_0$ to $Q_4$ are: less than 5, $[5,14)$, $[14,39)$, $[39,135)$, and greater than 135 received citations; and less than 3, $[3,5)$, $[5,8)$, $[8,17)$, and greater than 17 publications.
For example, a scientist who published seven papers and accumulated 100 citations would be assigned productivity impact group $Q_2$ and citation impact group $Q_3$.

\subsection*{Defining brokerage}
In the collaboration network, we define brokerage events as the formation of triangles between three authors $a$, $b$ and $c$.
For a given triplet of authors, we consider the unique ordering $(a,b,c)$, given by the temporal order of their initial pairwise collaborations, $t_{ab} < t_{bc} < t_{ac}$.
This strict order ensures that author $b$ has collaborated with both $a$ and $c$ before $a$ and $c$ collaborate for the first time together, resembling the definition of brokerage.~\cite{obstfeld_socialnetworkstertius_2005}
Note that at time $t_{ac}$, one out of two brokerage events may occur: one where all three authors co-author a publication together, and another where only authors $a$ and $c$ co-author a paper.
Special cases, e.g., when all three collaborate at the second timestamp ($t_{bc} = t_{ac}$), are not considered in the main analysis (see SI Note S2 for a descriptive analysis).

\subsection*{Brokerage's relationship to academic impact}
Using the closing collaboration timestamp $t_{ac}$, we collect the brokerage counts ${B}_m(s_i)$ for each author's impact group $Q_m$ at each career stage $s_i$.
To account for varying career lengths when comparing brokerage across stages, we define the brokerage rate $R_m(s_i) = B_m(s_i) / \Delta t_{s_i}$, representing the annual brokerage events during a given career stage.

As both the count and rate distributions are broad (SI Figs. S5 and S6), we evaluate the significance of our results using the Mann-Whitney test statistic~\cite{mann.whitney_testwhetherone_1947}
(see SI Fig. S7 and S8A for alternative metrics).
The reported statistic, known as the common language effect size,~\cite{mcgraw.wong_commonlanguageeffect_1992} estimates the probability that a randomly selected brokerage value (count or rate) from one group (career stage or impact group) is larger than a randomly selected brokerage value from another group.
The null hypothesis states that the distributions of brokerage of both groups are identical, and we reject it if the test statistic is significantly different from $50\%$ (the neutral line in~\cref{fig:03_bf_success,fig:04_bfr}).
We use permutation tests~\cite{ernst_permutationmethodsbasis_2004} to estimate the significance of the observed test statistic.
This test randomly assigns the brokerage frequencies (or rates) among the two groups and computes the Mann-Whitney test statistic for each random re-assignment.
A total of \Npermu~such re-samples yields the null distribution.
We report an observed difference as significant if it is more extreme than $95\%$ of the values in the null distribution.
Moreover, we estimate the $95\%$ confidence intervals for the test statistic using bootstrapping with \Npermu~re-samples of both groups.~\cite{efron.tibshirani_introductionbootstrap_1994}

When explicitly measuring the cumulative nature of brokerage (\cref{fig:04_bfr}B), we report the Spearman correlation coefficient between the brokerage rates of two consecutive career stages $s_i \leftrightarrow s_{i+1}$ of scientists in a given impact group.
In this context, we apply paired permutation tests to estimate the significance of the observed correlation.
This test randomly shuffles the brokerage rates within each career stage $s_i$ and $s_{i+1}$, breaking the observed pairing.
Again, we only report correlations at the $95\%$ significance level and bootstrap confidence intervals.

\subsection*{Gendered brokerage}
Given that an author may be known by multiple variations of their name, we apply a recently developed algorithm to infer the gender of each name (i.e., woman or man).~\cite{buskirk.etal_opensourceculturalconsensus_2023}
The final gender is then assigned to each author based on majority voting, ignoring the label ``unknown.''
A certainty threshold of $\ThreshGender$ (automatically inferred) leaves \Nauthunid~authors unidentified.
Of the remaining \Nauthid~scientists, only \Nauthidwoman~($\approx \Pauthidwoman$) are labeled as women.
We acknowledge that gender is a social construct and not binary; however, for this study, we focus on women and men due to the limitations of the inference algorithm we are using and because they collectively represent the majority of the data.
We leave the inclusion of non-binary or unisex names for future work.

To estimate the representation of women and men in a given year, we count the number of disambiguated authors who have their first publication before and their last publication after that year (\cref{fig:05_gender}A).
The decrease in the number of active authors in later years is likely due to falsely flagging authors as inactive who continue to publish after the observation period.
As this affects scientists of both genders equally, its impact on the comparison of normalized brokerage counts (SI Fig. S13) is expected to be small.
When quantifying the evolution of brokerage by gender composition, we focus on the 237,245 ($\approx$ 42$\%$) brokerage events $(a,b,c)$ for which we have identified the gender $g_a$, $g_b$ and $g_c$ of all three participants (\cref{fig:05_gender}B).
The reported brokerage event counts are summed up for the mixed-gender brokerage events.
For instance, the reported count of $(g_a, g_b, g_c) = (w,m,m)$ contains the sum of counts of the other brokerage events between one woman and two men, $(m,w,m)$ and $(m,m,w)$.

When reporting the probability densities of career stages (\cref{fig:05_gender}D--E), we consider the career stage of the three participants at the time of the brokerage event $t_{ac}$.
To obtain the probability density estimate (\cref{fig:05_gender}D), we normalize the counts in each stage by the duration of the stages, such that the sum over all stages is one.
For the joint distributions (inset of \cref{fig:05_gender}E), we consider the duration of both stages.
Finally, we compute the ratio between the joint seniority distributions of authors $a$ and $c$ of a given their genders and the overall population as $P_{g_a,g_c}(s_a,s_c)/P(s_a,s_c)$ (\cref{fig:05_gender}E).
Values above one indicate an over-representation of the specific genders in the given career stages in comparison to all brokerage events.
Values below one indicate under-representation and values equal to zero indicate no change.

%% file: 05_data_code_availability.tex
\section*{Data availability}
The APS datasets are available upon request to the American Physical Society (\href{https://journals.aps.org/datasets}{https://journals.aps.org/datasets}).
Intermediate results to reproduce the findings will be published at ~\href{https://zenodo.org/doi/10.5281/zenodo.12724812}{https://zenodo.org/doi/10.5281/zenodo.12724812} before the publication of this study.

\section*{Code availability}
The relevant scripts to reproduce our findings from the APS datasets are available at~\href{https://github.com/mannbach/cumulative\_advantage\_brokerage}{https://github.com/mannbach/\\cumulative\_advantage\_brokerage}.
The repository is linked to the data archive in which intermediate results will be made available (\href{https://zenodo.org/doi/10.5281/zenodo.12724812}{https://zenodo.org/doi/10.5281/zenodo.12724812}).
Before publication, the code will be extended to produce and import these intermediate results.

%% file: 06_acknowledgements.tex
\section*{Acknowledgements}
We thank Markus Strohmaier, Denis Helic, Roberta Sinatra, and the CSH Network Inequality group for their useful feedback on the manuscript, and Mingrong She for her help with the data curation.
J.B. was supported by the Austrian Science Promotion Agency FFG under project No. 873927 ESSENCSE.
J.B. is a recipient of a DOC Fellowship of the Austrian Academy of Sciences at the Complexity Science Hub.
G.I. acknowledges support from AFOSR (Grant No. FA8655-20-1-7020), project EU H2020 Humane AI-net (Grant No. 952026), and CHIST-ERA project SAI (Grant No. FWF I 5205-N).
L.E.N received support from the Vienna Science and Technology Fund WWTF under project No. ICT20-079.

\section*{Author contributions}
All authors conceptualized the study, wrote the manuscript and analyzed the results.
J.B. performed the computations, curated the data, visualized the results and wrote the software.
G.I. and J.B. designed the methodology to conduct the brokerage impact analysis; F.K., L.E.N. and J.B. conceived the methodology for the gender disparity analysis.
F.K. and G.I. supervised the execution and planning of the research activity.

\section*{Competing interests}
The authors declare no competing interests.

%% file: si_00_dataset.tex
\section{Dataset}
\label{ssec:s_aps}
The American Physical Society (APS) dataset contains the longitudinal record of papers published in associated journals between July 1st, 1893 and December 31st, 2020.
Earlier versions have been used in similar studies.~\cite{sinatra.etal_quantifyingevolutionindividual_2016,kong.etal_influencefirstmoveradvantage_2022}

\subsection*{Name disambiguation}
\label{sssec:s_name_disamb}
\begin{figure}[h]
    \centering
    \includegraphics[width=\textwidth]{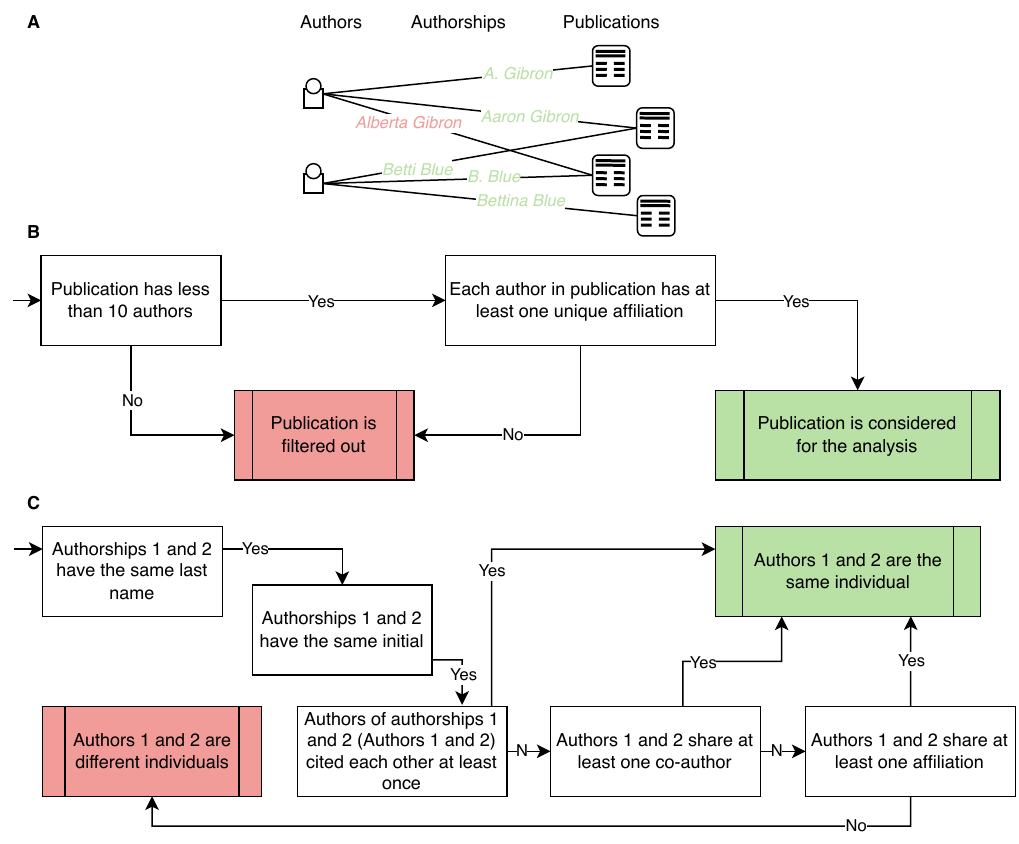}
    \caption[Name disambiguation algorithm schema]{\textbf{Name disambiguation algorithm schema} as proposed by Sinatra et al.~\cite{sinatra.etal_quantifyingevolutionindividual_2016}. The schema is inspired by a similar study, but extended.~\cite{kong.etal_influencefirstmoveradvantage_2022} (\textbf{A}) The goal of the algorithm is to infer the correct authorships, that is, the author identity for an author name given on a publication. (\textbf{B}) Filtering step removing publications with more than ten authors, as individual contributions are estimated to be less essential. (\textbf{C}) The conditions for two author names with identical last names and first name initials to be assigned the same identity. Their author identities are merged if they cite each other (self-citation), share a common co-author, or have a common affiliation. This solution is prone to assigning names with mismatching first names to the same author (red link in \textbf{A}).}
    \label{fig:s_name_disamb_exp}
\end{figure}

The original dataset does not contain author information beyond their name and affiliation specified on a given publication.
Tracking an author's contributions across multiple papers is non-trivial, given that names are ambiguous.
Distinct authors might share the same name and a single author might publish papers with variations of their name.
Identifying the correct links, or \textit{authorships}, between author identities and publications is a problem commonly known as author name disambiguation (example in~\cref{fig:s_name_disamb_exp}A).~\cite{tekles.bornmann_authornamedisambiguation_2020}

We build on an existing algorithm for the APS dataset that disambiguates authors based on their name, affiliation, shared co-authors and citation record.~\cite{sinatra.etal_quantifyingevolutionindividual_2016}
Two author names are merged under the same author identity if they share the same last name and their first names or initials match.
Moreover, they either have to share a common co-author, cite each other at least once or have a similar affiliation.
If these conditions are met, the two author names and their publication contributions are assigned to the same author identity.
The definition of this algorithm is ambiguous on how to merge authors who already have multiple first names assigned to them.
Hence, we first base our re-implementation on a schema presented in a subsequent study that merges author identities if their first initials match (see~\cref{fig:s_name_disamb_exp}B for an adjusted schematic depiction).~\cite{kong.etal_influencefirstmoveradvantage_2022}

\begin{figure}[h]
    \centering
    \includegraphics[width=\textwidth]{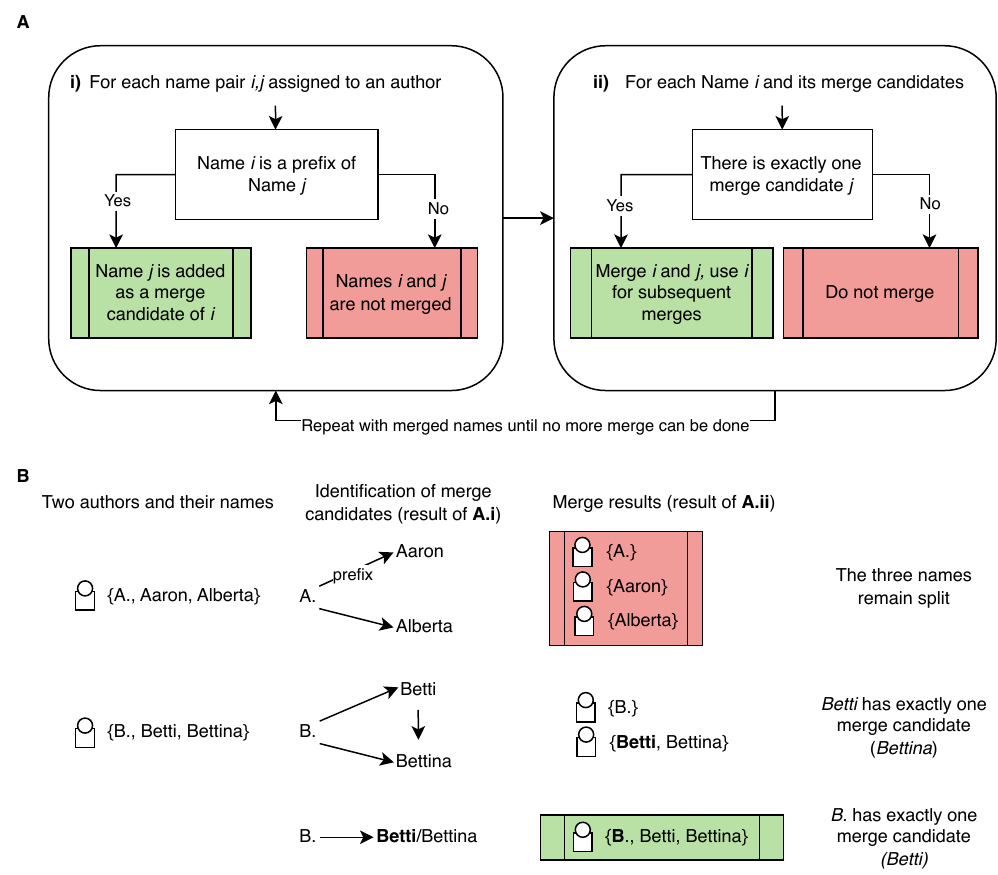}
    \caption[Name disambiguation extension schema]{\textbf{Name disambiguation extension schema.} (\textbf{A}) Multiple names assigned to the same author as a result of the name disambiguation by Sinatra et al.~\cite{sinatra.etal_quantifyingevolutionindividual_2016} (as explained in~\cref{fig:s_name_disamb_exp}) are re-assessed in two steps: (i) For a name $i$, we collect as merge candidates all first names $j$ assigned to the same author, if $i$ is a prefix of $j$. (ii) For all names with unique candidates, we merge the corresponding author identities. We repeat these steps until no merges can be done anymore. (\textbf{B}) Two examples of the name disambiguation algorithm extension. The first example fails, because \textit{A.} is a prefix both to \textit{Aaron} and \textit{Alberta}. We divide the names into three identities and assign the respective publications.  The second example succeeds in two steps, as \textit{Betti} is a prefix only to \textit{Bettina}. After merging the two, \textit{B.} is a sole prefix of \text{Betti/Bettina}. The names remain assigned to the same identity, including their respective publications.}
    \label{fig:s_name_disamb_ext}
\end{figure}

\begin{figure}[h]
    \centering
    \includegraphics[width=\textwidth]{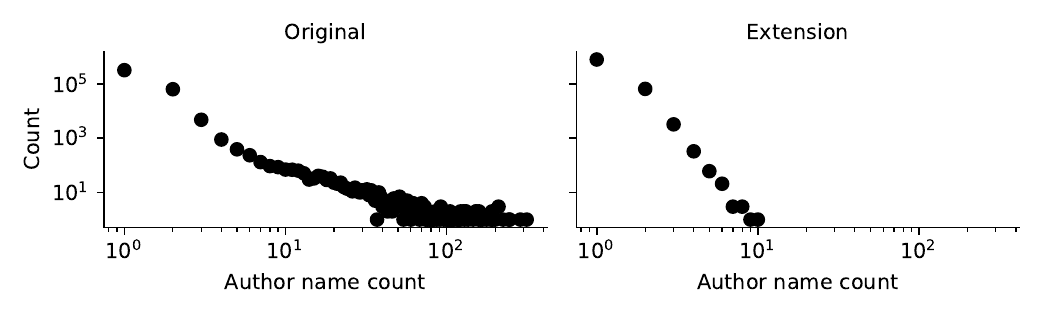}
    \caption[Author names per author frequency distribution]{\textbf{Author names per author frequency distribution.}
        Frequencies for the number of author names assigned to a single author identity before (left column) and after (right column) applying the name disambiguation extension (\cref{fig:s_name_disamb_ext}).
        The extension reduces the maximum number of assigned names to a single author from over three hundred to ten, following a continuous decrease in the probability of name counts.}
    \label{fig:s_data_disamb}
\end{figure}

This solution is prone to falsely merging authors with first names mismatching beyond the first initial, thereby creating authors with more than 300 name variations assigned to them (left column in~\cref{fig:s_data_disamb}).
These errors are problematic as they inflate the activity of the falsely merged author, expanding their local network and potentially creating false brokerage events.
To reduce this merging error, we implement a post-processing algorithm that is stricter when deciding which names to assign to the same author identity (\cref{fig:s_name_disamb_ext}).
In summary, the algorithm splits all first names that were previously assigned the same author if they are not unique prefixes of one another.
It iteratively tries to merge name pairs back together if a first name is a prefix to exactly one other first name.
This process is repeated until no more merges can be made.
Names that do not have a unique solution are kept separate, e.g., \textit{A.}, \textit{Aaron} and \textit{Alberta}.

The resulting distribution of author name counts per individual implies a more realistic assignment (right column in~\cref{fig:s_data_disamb}).
The probability of an author name count decreases consistently, with ten being the highest observed count of author names assigned to a single author.
As this procedure only splits previously merged names, it can only decrease the false merge error.
It could, however, result in an increase in the complementary error type by falsely removing authorships that belong to the same person.
For instance, a name triplet of (\textit{B.}, \textit{Betty} and \textit{Bettina}) would be split into three author identities as \textit{B.} is a prefix of both other names and \textit{Betty} cannot be merged into \textit{Bettina}.
The publications of either \textit{Betty} or \textit{Bettina} published under their initial \textit{B.}, can not be assigned to them.

To quantify this trade-off, we sample 50 pairs of papers in which the inferred authorships of the two algorithms differ.
These are authorships that the first solution assigns to the same author, while the extension assigns two separate authors.
For each such pair, we search for a link between both papers under the same author identity.
We consider the sampled pair if we find an online CV, Google Scholar profile, ORCID profile, or a research gate profile or list of publications from a university or personal website listing at least one of the two articles, considering the respective author names.
If both papers appear in such a list we label the two pairs to belong to the same person (in agreement with the first solution).
If only one publication is listed, we count it as an agreement with the extension algorithm.
Through this procedure, we find that 32 out of 44 pairs were correctly split.
Hence, the improved solution does not only provide a more conservative solution (preferring split identities over merged ones) but also appears to be more accurate.

\subsection*{Gender inference}
To guess the most probable gender assigned to a given author name, we employ a recently proposed solution.~\cite{buskirk.etal_opensourceculturalconsensus_2023}
As compared to existing, mostly proprietary, services, the authors have made both the comprehensive source datasets and inference model publicly available.\footnote{\url{https://github.com/ianvanbuskirk/nomquamgender}}
The final assignment is based on a weighted average across 36 data sources, returning the probability that a given name $m$ is assigned a female label $p_{gf}^m$.
A value close to $p_{gf}^m \approx 0.0$ indicates that the given name is strongly gendered male, whereas $p_{gf}^m \approx 0.5$ expresses uncertainty across the source data.
The latter could be due to the name being unisex or spatial or temporal dependencies.
To assign a label to an author name $m$ we choose a threshold of $34\%$, meaning that names with $0.34 < p_{gf}^m < 0.66$ remain unclassified.

%% file: si_01_intermediate_brokerage.tex
\section{Intermediate brokerage}
\label{ssec:s_brokerage}
We report the number of cases in which three authors collaborate in a triangle.
When enforcing the temporal order $t_{ab} < t_{bc} < t_{ac}$, we identify a total of 567,201 brokerage events, as reported in the main text.
Partially relaxing the temporal order by allowing $t_{ab} = t_{bc} < t_{ac}$, we find 1,284 additional cases.
In this scenario, authors $a,b$ and $b,c$ publish their first article at the same time (but not all together), before $a$ and $c$ collaborate at a later point in time.
Much more common, however, is the case where $t_{ab} < t_{bc} = t_{ac}$, which results in 1,240,087 additional cases.
In this case, $a$ and $b$ collaborate first, before $b,c$ and $a,c$ publish together at a second point in time.
This case is prominent as it includes the scenario where $a$, $b$ and $c$ publish an article together at time $t_{bc} = t_{ac}$.
It resembles a repeated collaboration between $a$ and $b$ with a new collaborator $c$.
Even more probable is the formation of an immediate triangle among three authors who have never collaborated before, i.e. $t_{ab} = t_{bc} = t_{ac}$, resulting in 1,417,005 additional cases.
We focus our analysis on the strict temporal order $t_{ab} < t_{bc} < t_{ac}$ as it ensures that the broker $b$ has collaborated with both $a$ and $c$ before they collaborate.

%% file: si_02_careers.tex
\section{Measuring brokerage throughout careers}
\begin{figure}[h]
    \centering
    \includegraphics[width=0.6667\textwidth]{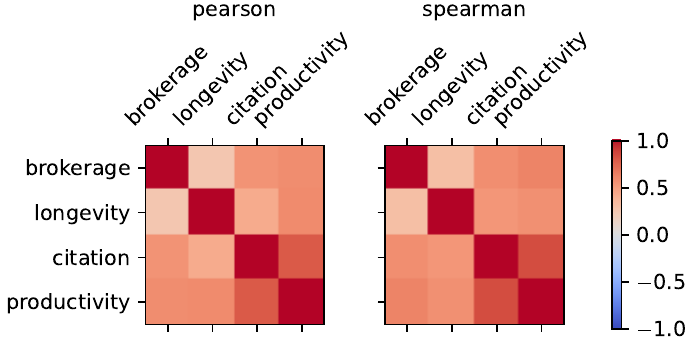}
    \caption[Correlation matrix of impact, career length and cumulative brokerage]{\textbf{Correlation matrix of impact, career length and cumulative brokerage} at the end of a career using the Pearson (left) and Spearman (right) correlation coefficient. The final cumulative brokerage frequency correlates with the number of publications, citations, and career length.}
    \label{fig:si_corr_heat}
\end{figure}

\begin{figure}[h]
    \centering
    \includegraphics[width=\textwidth]{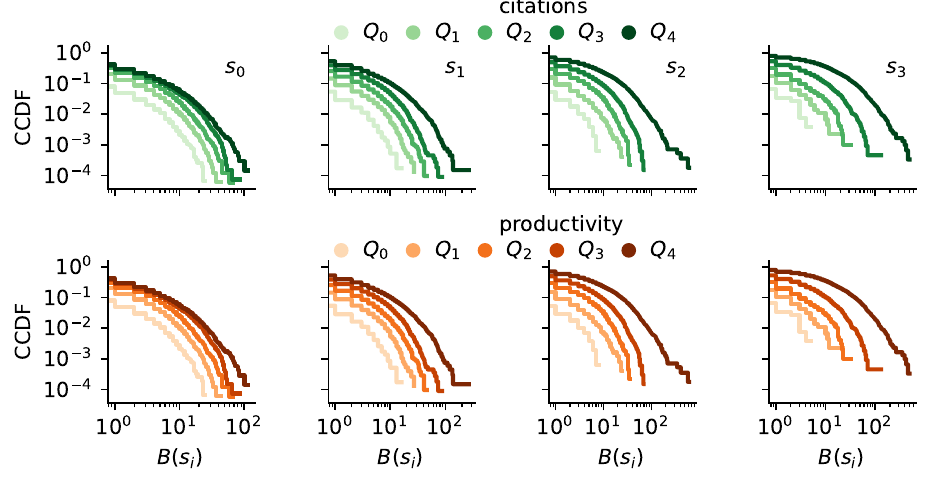}
    \caption[Brokerage frequency distributions]{
        \textbf{Brokerage frequency distributions} across career stages (columns) and impact groups (color intensity) of citation and productivity impact groups (upper and lower row, respectively). Brokerage frequency $B(s)$ values are aligned column-wise to match the varying participation throughout careers. The distributions show the qualitative differences quantified and summarized in the main text. The more successful groups tend to have higher brokerage frequencies and the effect grows with career stages. The distributions are broad, which could make a comparison by means sensitive to extreme values.}
        \label{fig:si_bf_ccdf}
\end{figure}

\begin{figure}[h]
    \centering
    \includegraphics[width=\textwidth]{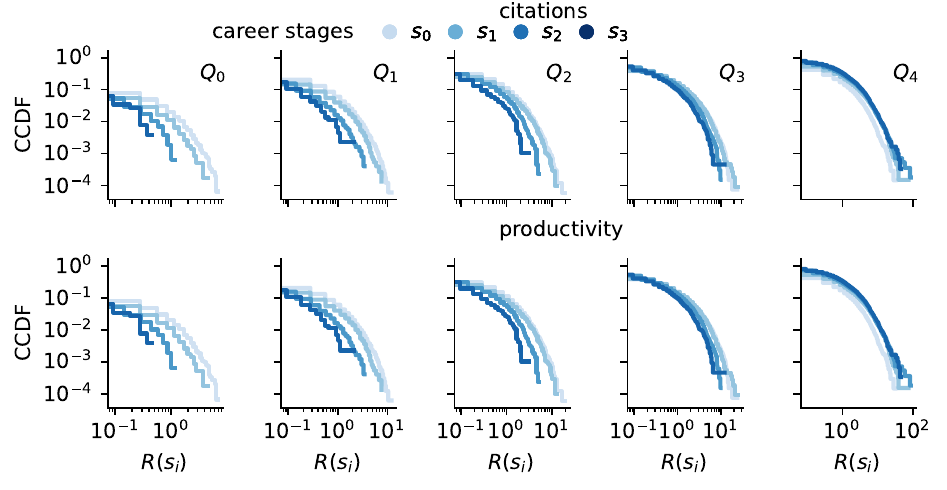}
    \caption[Brokerage rate distributions]{
        \textbf{Brokerage rate distributions} across success groups (columns) and stages (color intensity) of citation and productivity (upper and lower row, respectively). Brokerage rates $R(s)$ values are aligned column-wise to match the varying participation throughout success groups. We observe the transition from slow-down to speed-up in brokerage rates across increasingly successful careers and stages. Moreover, the distributions are broad, which could make a comparison by means sensitive to extreme values.}
    \label{fig:si_br_ccdf}
\end{figure}

%% file: si_03_metrics.tex
\section{Alternative metrics and covariates measuring the impact of brokerage}
\begin{figure}[h]
    \centering
    \includegraphics[width=\textwidth]{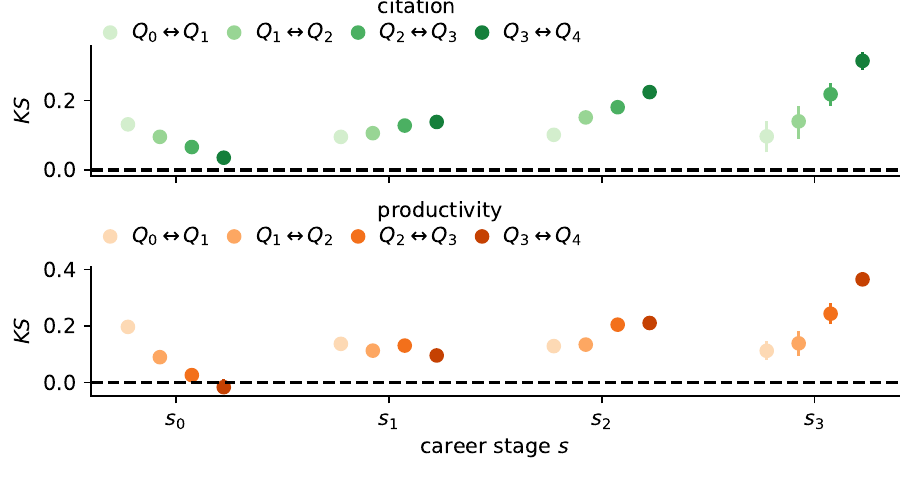}
    \caption[Brokerage frequency comparison using Kolmogorov-Smirnov test statistic]{\textbf{Brokerage frequency comparison using the Kolmogorov-Smirnov test statistic.} The Kolmogorov-Smirnov test provides an alternative to the Mann-Whitney test for comparing brokerage frequencies between impact groups. The test statistic (KS) is the maximum vertical distance between the cumulative distribution functions of two samples. The KS test statistic is zero if the two samples are drawn from the same distribution. Values above zero indicate that the more successful group tends to have higher brokerage frequencies. The results qualitatively match the Mann-Whitney test results (main text Fig. 3).}
    \label{fig:si_bf_success_ks}
\end{figure}

\begin{figure}[h]
    \centering
    \includegraphics[width=\textwidth]{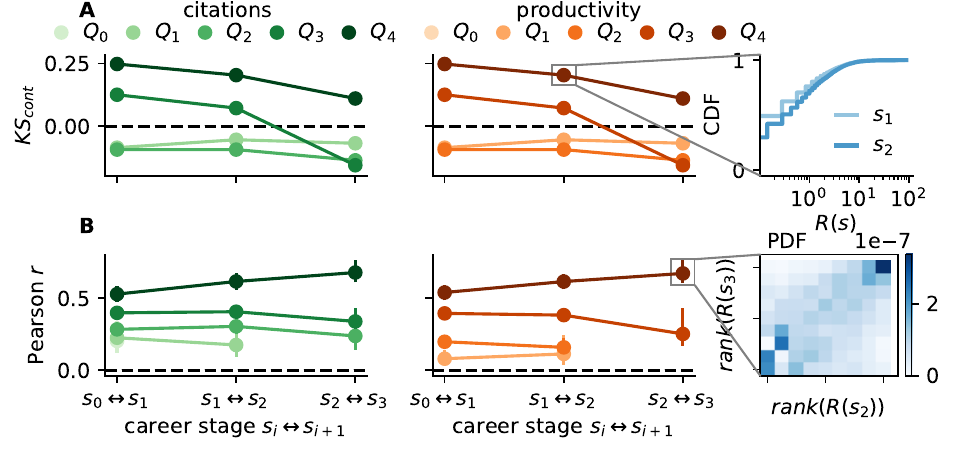}
    \caption[Alternative brokerage rate comparison and correlation]{\textbf{Alternative brokerage rate comparison and correlation.} The results qualitatively match the brokerage rate comparisons presented in the main text (Fig. 4). (\textbf{A}) The continuous Kolmogorov-Smirnov test statistic ($KS_{cont}$) is the maximum vertical distance between the cumulative distribution functions of two samples. The KS test statistic is zero if the two samples are drawn from the same distribution. Values above zero indicate that the more successful group tends to have higher brokerage rates in a given career stage. (\textbf{B}) The Pearson correlation coefficient quantifies a linear relationship between two consecutive brokerage rates of a single scientist. Similar to the results of the Spearman correlation coefficient, it is positive and significant, indicating that scientists who participate in brokerage during one career stage are likely to continue doing so in the next stage. The effect is strongest for the most successful scientists.}
    \label{fig:si_br_comp_ks_pearson}
\end{figure}

\begin{figure}[h]
    \centering
    \includegraphics[width=\textwidth]{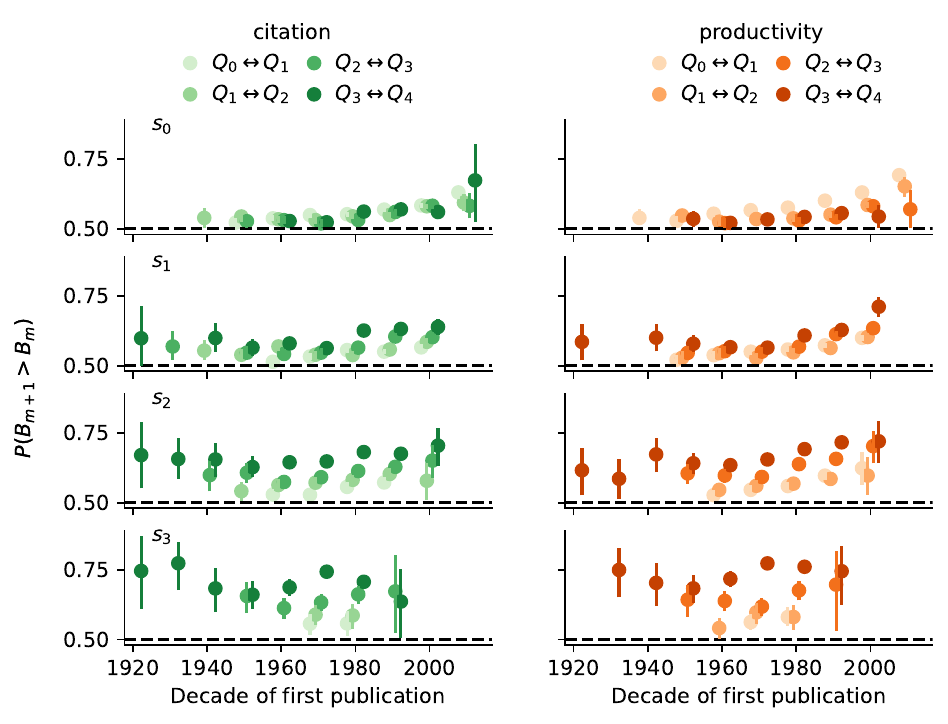}
    \caption[Brokerage frequency comparison per cohort decade]{
        \textbf{Brokerage frequency comparison per cohort decade.} Comparison of brokerage participation across citation and productivity impact groups (columns, increasing color intensity for higher impact comparisons), within stages (rows) and cohorts starting in the same decade (x-axis). Brokerage is likely to co-evolve with other longitudinal academic processes, such as the growing importance of collaboration teams~\cite{wuchty.etal_increasingdominanceteams_2007} or the rapid growth of publication counts~\cite{fire.guestrin_overoptimizationacademicpublishing_2019}. When partitioning careers into stages, we might bias the results towards more recent versus scientists of prior decades. We account for this by grouping scientists based on the decade they started publishing in, and compare them only within their cohort. Our main results hold for most decades. The effect of brokerage is generally positive (all markers above the neutral line) and the effect grows with impact and career stages. Results in early or late cohorts are more noisy due to the smaller sample sizes.}
    \label{fig:si_decade_bf_cmp}
\end{figure}

%% file: si_04_gender_impact.tex
\section{Brokerage impact by gender}

\begin{figure}[h]
    \centering
    \includegraphics[width=\textwidth]{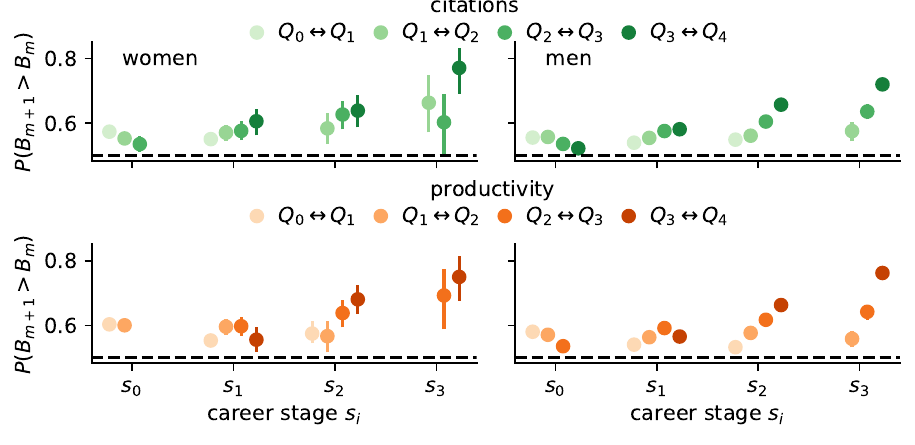}
    \caption[Brokerage frequency comparison per gender]{\textbf{Brokerage frequency comparison per gender.} Comparing brokerage frequencies between consecutive impact groups $Q_m \leftrightarrow Q_{m+1}$ in a given career stage $s_i$ and within a gender group (columns). Splitting the sample of scientist by their inferred gender and comparing only within groups accounts for the gender disparities in academia~\cite{huang.etal_historicalcomparisongender_2020}. Brokerage is equally beneficial for women and men.}
    \label{fig:si_bf_cmp_gender}
\end{figure}

\begin{figure}[h]
    \centering
    \includegraphics[width=\textwidth]{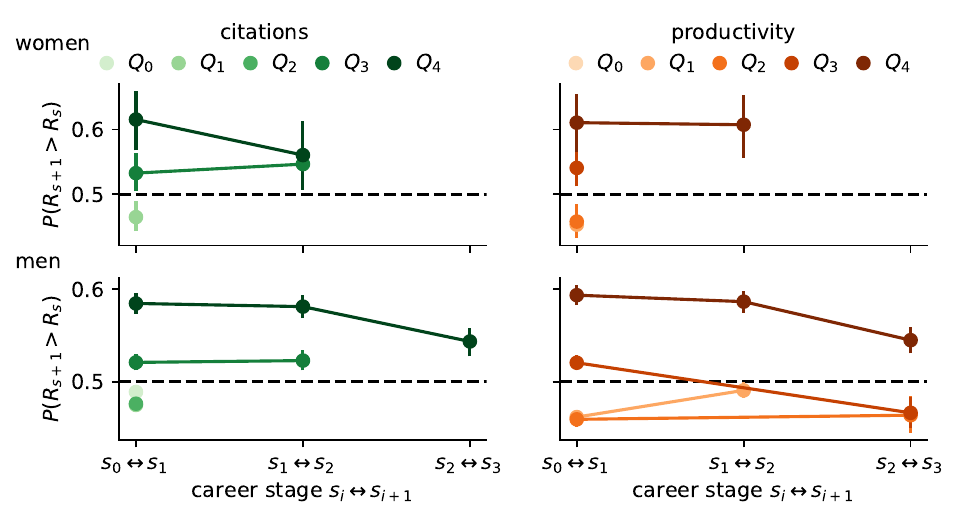}
    \caption[Brokerage rate comparison per gender]{\textbf{Brokerage rate comparison per gender.} Comparing brokerage rates between consecutive career stages $s_{i} \leftrightarrow s_{i+1}$ for a given impact group $Q_m$ and within a gender group (rows). Among both women and men, only the most successful group has a consistent speed-up of brokerage throughout their careers. The effect is identical for women and men. As an effect of women's smaller group size, their results are more noisy.}
    \label{fig:si_br_cmp_gender}
\end{figure}

\begin{figure}[h]
    \centering
    \includegraphics[width=\textwidth]{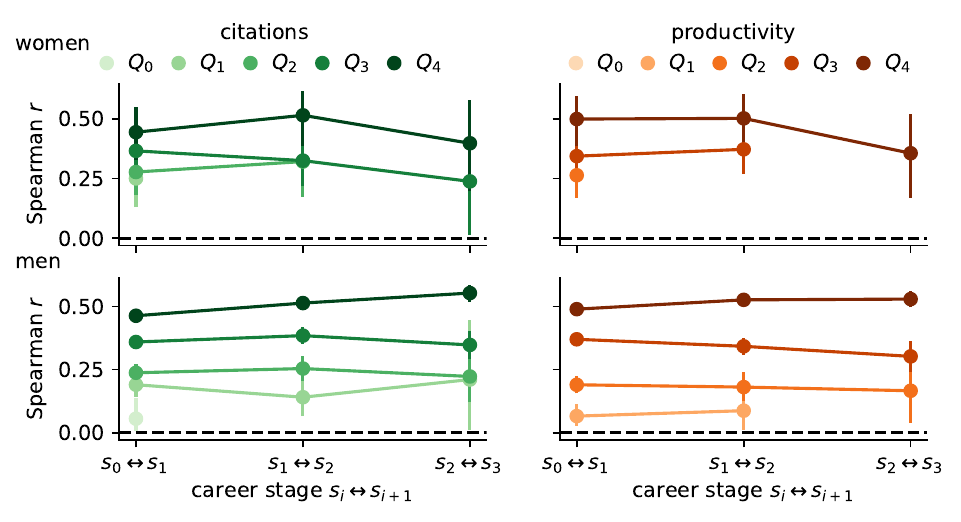}
    \caption[Brokerage rate correlation per gender]{\textbf{Brokerage rate correlation per gender.} Correlating brokerage rates between consecutive career stages $s_{i} \leftrightarrow s_{i+1}$ for a given impact group $Q_m$ and within a gender group (rows). Although we observe slightly more variation for women, brokerage is cumulative for scientists of both genders, increasingly so for the more successful groups.}
    \label{fig:si_br_cor_gender}
\end{figure}

%% file: si_05_normalization.tex
\section{Normalizing gendered brokerage events by group sizes}
\label{ssec:s_brokerage_evolution_gender}
We report the annual number of brokerage events normalized by the total number of possible triangles among active authors in the respective year (\cref{fig:si_decade_bf_cmp}).
We consider authors as active in a given year if their first publication record is in that year or before and they have their last recorded publication in a later year.
By reporting the normalized counts until 2016, we reduce the error of falsely marking authors as inactive.
For the normalization of the brokerage event count among $k_m$ male and $k_w$ female scientists, we compute the number of possible triangles as
\begin{equation}
    \binom{N_m(y)}{k_m}\times \binom{N_w(y)}{k_w}
\end{equation}
where $N_m(y)$ and $N_w(y)$ are the number of active male and female scientists in a given year $y$.
Note that we have $k_w + k_m = 3$ and $0 \leq k_m,k_w \leq 3$ because always three authors engage in a single brokerage event.
For instance, to normalize the count of brokerage events between two women and one man, we use $k_m=1$ and $k_w=2$.

\begin{figure}[h]
    \centering
    \includegraphics[width=\textwidth]{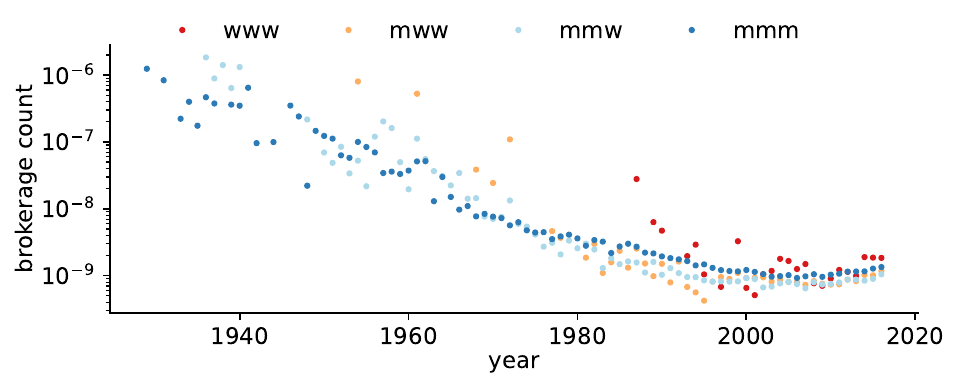}
    \caption[Normalized gender brokerage counts]{
        \textbf{Normalized gender brokerage counts}. The annual brokerage event counts per gender composition (color) are normalized by the total number of possible triangles among active authors in the respective year (see~\cref{ssec:s_brokerage_evolution_gender}), estimating the fraction of brokerage events from what is possible. The counts of different gender compositions mostly follow the same trend, indicating that the group size disparities alone explain the differences in brokerage participation.
    }
    \label{fig:si_brok_cnt_gender_norm}
\end{figure}